# On the Refractive Index of Ageing Dispersions of Laponite

N. V. N. Ravi Kumar†, K. Muralidhar‡, Yogesh M. Joshi†, *

†Department of Chemical Engineering

‡Department of Mechanical Engineering and Centre for Laser Technology

Indian Institute of Technology Kanpur, Kanpur 208016 INDIA

* Corresponding Author, E-Mail: joshi@iitk.ac.in,

Telephone: +91 512 259 7993, Fax: +91 512 259 0104

**Abstract**

Aqueous dispersion of Laponite at low ionic concentration is of interest since it undergoes structural evolution with respect to time, which is usually termed as ageing. In this work we study the refractive index behavior as a function of ageing time, concentration and temperature. We observed that the extended Lorenz-Lorentz equation fitted the refractive index dependence on concentration and temperature very well. The refractive index did not show any dependence on ageing time. However, the dependence of refractive index on concentration showed a marked change as the system underwent transition from an isotropic to a biphasic state. The slope of the refractive index-density data is remarkably close to that of water at all Laponite concentrations. In the context of transport phenomena, optical measurements such as interferometry can exploit the water-like behavior of Laponite dispersions.

# 1. Introduction

A variety of complex fluids of technological importance such as pastes, gels, concentrated dispersions and emulsions and other soft materials fail to attain equilibrium over practical time scales. Their structure continuously evolves with time and is accompanied by a significant change in various physical properties such as shear modulus and viscosity. Such soft glassy materials, also known as nonergodic materials, show extremely slow relaxation behavior, with relaxation time scaling with the ageing time [Cipelletti and Ramos, 2005]. Some of the prominent systems that show nonergodic behavior are aqueous dispersions of smectites [Lagaly, 1989]. In the last decade, significant work has been carried out to understand ageing and related glassy behavior of these dispersions [Andreola et al., 2006; Bonn et al., 1999; Schosseler et al., 2006, Joshi, 2007; Joshi et al., 2008; Joshi and Reddy 2007]. An aqueous dispersion of Laponite, is also known to undergo ergodic to non-ergodic transition over measurable time scales and has attracted significant attention due to its unusual physical behavior. A property not adequately addressed is refractive index of such dispersions. In this paper, we study refractive index dependence of aqueous Laponite dispersion on concentration, ageing time and temperature.

Laponite has many important industrial applications ranging from agriculture, construction, personal care, surface coatings, paper, and polymer industry. Apart from industrial applications, Laponite-in-water is considered to be a model system to study phenomena such as glass transition and gelation [Cipelletti and Ramos, 2005]. Laponite particles have a disk-like shape with a thickness around 1 nm and a diameter of 25 nm. In a salt-free aqueous medium, the negative charge on the face leads to electrostatic screening length of 30 nm on its surface [Bonn et al., 1999]. The aqueous dispersion results in the formation of non-ergodic soft solids at very low concentration. This is because the effective volume associated with each disc, the Debye-Huckel sphere, is much larger than the disc itself [Mossa et al., 2007].

As an aqueous Laponite dispersion undergoes structural evolution and ageing, viscosity and elastic modulus increase by several orders of magnitude [Bonn et al., 1999]. It is observed that lower the concentration, slower is the increase in viscosity [Abou et al., 2001]. Furthermore, ageing of the dispersion is accompanied by a faster increase in viscosity and elastic modulus at higher temperatures [Ramsay, 1986]. The ageing properties of the system have been extensively characterized by various optical techniques in the literature.

In the present study, the dependence of refractive index of aqueous Laponite dispersion at low ionic concentration is investigated with respect to concentration of Laponite, ageing time and temperature. The extent to which the data can be fitted by an extended Lorenz-Lorentz equation is examined. Laponite-in-water is a subject of research that often employs optical characterization. This is the first study that investigates the effect of system variables on refractive index of ageing colloidal glass of clay mineral dispersions. The refractive index behavior reported in the present work will be useful in understanding the physical phenomena that underlie age-related transition.

2. Experimental

Laponite RD, synthetic hectorite clay was procured from Southern Clay Products, Inc. Laponite as white powder was dried for 4 hours at 120°C before mixing with ultra pure water at pH 10 under vigorous stirring. Basic pH, maintained by the addition of NaOH, provides chemical stability to the dispersion [Mourchid and Levitz, 1998]. The dispersion was stirred vigorously for 1 hour by Ultra turrax T25 at a speed of 9,500 rpm. The sample used to carry out RI measurement was filtered through a Millipore Millex-HV 0.45 μm filter unit. Concentrations ranging from 2 weight % to 4 weight % and temperatures from 15°C to 40°C have been used in the study. Refractive index measurements were carried out by a digital refractometer (Atago RX5000α) [Niskanen et al., 2006]. The light source used in the unit is LED (with a strong unpolarized D line at a wavelength of 589.3 nm). The refractometer allowed measurement of refractive index with a precision of $\pm 4 \times 10^{-5}$. Before making measurements, the refractometer was calibrated and standardized using a procedure suggested by the manufacturer. The measurement was subjected to a repeatability test and the uncertainty in the refractive index was found to be ±0.004%.

## 3. Results

Figure 1 shows refractive index (RI) data at 20°C for two concentrations of 3 weight % and 3.5 weight % with respect to ageing time. Ageing time is measured as time elapsed after carrying out filtration of the sample into the refractometer. Figure 1 shows that RI does not change with respect to ageing time for up to 7 hours in the 3.5 weight % Laponite dispersion. In addition, we did not observe any noticeable difference in RI of the unfiltered sample in comparison with the filtered sample. Over this time period, Laponite dispersion is known to undergo ergodicity breaking [Schosseler et al., 2006] and a change in viscosity of over 4 orders of magnitude [Bonn et al., 1999]. The observed behavior is in contrast with ageing polymeric glasses, where RI increases with the ageing time [Huang and Paul (2006); Tanio and Nakanishi (2006)].

Since ageing time does not have any influence on the RI of the Laponite dispersion (Figure 1), RI was measured with respect to concentration and temperature, irrespective of ageing time. In each measurement, it was ensured that steady state temperature was reached throughout the sample. Figure 2 shows the dependence of RI on temperature for various concentrations, including zero concentration, namely pure water. For a given concentration RI decreases with temperature with a dependence similar to that of water. The inset in Figure 2 shows that the rate of increase of RI with concentration for a given temperature is constant up to 3.5 weight % and decreases significantly at higher concentrations.

The refractive index variation of an isotropic single phase system with respect to density, temperature and wavelength of light is well represented by the Lorenz-Lorentz equation and is given by [Lorentz, 1952]:

$$\frac{n^2-1}{n^2+2}\frac{1}{\rho} = \frac{N_A}{3\varepsilon_0 M}\alpha(\rho,T,\lambda) \qquad (1)$$

Here, $n$ is refractive index, $\rho$ is specific gravity of the system, $N_A$ is Avogadro number, $\varepsilon_0$ is relative permeability, $M$ is molecular weight, $\lambda$ is wavelength of light, and $T$ is absolute temperature. Parameter $\alpha$ is mean polarizability and is found to be a function of temperature, density and wavelength of light [Tan, 1998]. Equation 1 holds for transparent isotropic media in which an inhomogeneous distribution of density prevails. Equation 1 can have a limited applicability to certain classes of anisotropic media as well [Wu, 1986]. Without loss of generality, the function on the right hand side of Equation 1 can be represented by a virial expansion of the arguments with undetermined coefficients. The wavelength of light was a constant in experiments. For the present system of Laponite dispersion, we utilize the first order virial expansion of the right hand side of Equation 1 leading to [Schiebener, et al., 1990],

$$\frac{n_D^2 - 1}{n_D^2 + 2} \frac{1}{\rho} = a_0 + a_1 \rho + a_2 T . \qquad (2)$$

Here, $n_D$ is the refractive index of the sample at the D line of the light source. In order to check validity of the above equation to Laponite dispersion, the density of the dispersion was corrected by including the concentration of Laponite (specific gravity of 2.5). Furthermore, the temperature dependence of density of the dispersion was estimated using the temperature dependence of density of water. Since the concentration of Laponite in experiments is very low, the temperature dependence of density of Laponite was neglected.

Figure 3 shows the composite quantity $(n_D^2 - 1)/\rho(n_D^2 + 2)$ computed using the measured RI data for various concentrations and temperatures and plotted as a function of the specific gravity of the dispersion. The lines show a best fit of Equation (2) to the experimental data with $a_0 = 0.3475$, $a_1 = -0.13$ and $a_2 = 4 \times 10^{-5} \text{K}^{-1}$. The fitting procedure employed is given below. We fitted a straight line to $(n_D^2 - 1)/\rho(n_D^2 + 2)$ vs. $\rho$ data at constant temperature. The fit yielded the slope to be $a_1$ and intercept to be $a_0 + a_2 T$. Parameter $a_1$ was observed to be practically independent of temperature. Finally, fitting the intercept vs. temperature data as a straight line yielded constants $a_0$ and $a_2$. It can be seen that Equation 2 provides a good fit to the experimental data, demonstrating the applicability of Lorenz-Lorentz equation to the ageing Laponite dispersion up to concentration 3.5 weight %. Beyond this concentration, a slight deviation from the Lorenz-Lorentz equation is apparent.

The derivative of $n_D$ with respect to density can be obtained as

$$\left(\frac{\partial n_D}{\partial \rho}\right)_T = \frac{\left(n_D^2 + 2\right)^2}{6n_D}\left(a_0 + 2a_1\rho + a_2 T\right). \tag{3}$$

For a weight % of 3.5 and a temperature of 20°C, the derivative is $1.68 \times 10^{-4}$ m$^3$/kg. The derivative of the Laponite dispersion can be compared with $1.78 \times 10^{-4}$ m$^3$/kg, the refractive index-density derivative of pure water. The two values are quite close to each other. This observation can simplify a family of measurement techniques such as optical interferometry [Goldstein, 1996].

## 4. Discussion

The trends seen in the refractive index behavior can be related to the microscopic structural changes that occur in the ageing system due to activated dynamics. In the powder form, Laponite discs are present in parallel stacks. Upon dispersing in water, the clay swells in two steps [Van Olphen, 1977]. In the first step, water is absorbed in successive monolayers on the surface that pushes the discs apart, while in the second stage of swelling, double layer repulsion pushes the Laponite discs farther apart. In general, the aqueous Laponite dispersion follows a two-stage ageing process. The first stage is termed as cage formation regime while the second is termed the full ageing regime [Tanaka et al., 2005]. Recently, Joshi (2007) proposed that after dispersing Laponite powder in water, clusters of Laponite undergo osmotic swelling with elapsed time (age). As various growing clusters meet, ergodicity breakdown occurs. In the full ageing regime, individual Laponite particles can be considered to be trapped in an energy well created by the surrounding particles. Mossa *et al.* (2007) observed that anisotropy of the platelet along with the repulsive potential is responsible for the local disorder that takes the system into the metastable state. The authors further observed that ageing dynamics strongly affects orientational degree of freedom that relaxes over the timescale of the translational modes. Thus, in the full ageing regime each particle undergoes a structural rearrangement with ageing time and attains a lower energy state. The physical state in the full ageing regime, whether glass or gel is a subject of intense research [Tanaka et al. (2004), Ruzicka et al. (2004), Joshi et al. (2008)]. The measurement of refractive index with respect to ageing time, shown in Figure 1, thus involves both cage forming as well as full ageing regime.

In the cage forming regime osmotic swelling of clusters occurs, though the system is still ergodic in nature. In the full ageing regime, ageing involves restructuring of the jamming entities by an activation process [Fielding et al., 2000] and the system is non-ergodic for length-scales greater than that of the colloidal object. This explains why in colloidal glasses, ageing does not involve volume relaxation and hence any change in density. Since the refractive index of a transparent medium is primarily dependent on density and temperature, a change in RI with respect to ageing time is thus not observed. However, for molecular glasses, volume relaxation (densification), that accompanies ageing increases the refractive index [Huang and Paul (2006); Tanio and Nakanishi (2006), Tan (1998)].

At Laponite concentrations of around 3.5 weight %, the dispersion is expected to undergo an isotropic to biphasic phase transition. In the biphasic state isotropic and nematic phases coexist with each other. This result can be obtained as follows. Laponite has an aspect ratio of $a=1/25$. Using Onsager's approach [de Gennes and Prost, 1993], the volume fraction at which discotic system undergoes isotropic to biphasic and biphasic to nematic transition are given by $\phi_{i-b}=0.33a$ and $\phi_{b-n}=0.45a$ respectively. Thus, the weight fraction at which isotropic to biphasic transition occurs is around 0.033, namely 3.3 weight %, corresponding to a volume fraction of 0.0133. Interestingly, based on the observations of Mourchid et al. (1995), in the phase diagram proposed by Ruzicka *et al.* (2006), the concentration regime above 3 to 3.5 wt. % has been termed as nematic gel, where local order exists among Laponite particles. This observation nicely matches with our conjecture. The refractive index data in Figure 2 for concentrations of around 3.5 weight % show the numbers getting saturated, though the trend with respect to temperature is entirely followed. Thus, refractive index clearly indicates isotropic to biphasic transition in the Laponite dispersion. The measured refractive index at concentrations greater than 3.5 weight % is probably an equivalent value of the biphasic structure, suitable for an anisotropic medium. However, under anisotropic conditions, it is appropriate to measure the directional components of the refractive index and is beyond the scope of the present work. Since Lorenz-Lorentz equation is applicable to only to isotropic systems, deviation of the experimental data for aqueous dispersions from the trend of Lorenz-Lorentz equation is justified. However, in general, the applicability of the extended Lorenz-Lorentz equation to an ageing dispersion of Laponite clay in the isotropic state is itself a significant result.

**5. Conclusions**

We experimentally studied the refractive index behavior of an aqueous dispersion of Laponite with respect to ageing time, temperature and concentration. Results show that refractive index does not change with the ageing time. Since refractive index is strongly dependent on density, the present study shows that volume relaxation phenomenon is absent in the Laponite dispersion, and in general, ageing colloidal glasses. These trends are in contrast to ageing molecular glasses that show a definite increase in density. The measured data fits well the extended Lorenz-Lorentz equation for the concentration and temperature dependence of refractive index. The slope of the refractive index-density line of the Laponite dispersion is close to that of pure water. As the concentration of Laponite in water increases, refractive index dependence of concentration shows diminished sensitivity, though the slope with respect to temperature remains unchanged. The change in sensitivity of the solution can be associated with transition from an isotropic to a biphasic phase, where the nematic phase coexists with the isotropic phase.

**Acknowledgement**: This work was supported by the BRNS young scientist research project awarded by Department of Atomic Energy, Mumbai to YMJ.

**References:**

Abou, B., Bonn, D., Meunier, J., 2001. Aging dynamics in a colloidal glass. Phys. Rev. E 64, 215101-215106.

Andreola, F., Castellini, E., Ferreira, J. M. F., Olhero, S., Romagnoli, M., 2006. Effect of sodium hexametaphosphate and ageing on the rheological behaviour of kaolin dispersions. Appl. Clay Sci. 31, 56-64.

Bonn, D., Kellay, H., Tanaka, H., Wegdam, G., Meunier, J. 1999. Laponite: What Is the Difference between a Gel and a Glass? Langmuir 15, 7534-7536.

Cipelletti, L., Ramos, L., 2005. Slow dynamics in glasses, gels and foams. J. Phys. Cond. Mat. 17, R253–R285.

de Gennes, P. G., Prost, J., 1993. The Physics of Liquid Crystals. Oxford, New York.

Fielding, S. M., Sollich, P., Cates, M. E., 2000. Aging and rheology in soft materials. J. Rheol. 44, 323-369.

Goldstein, R. J.1996. Fluid Mechanics Measurements. second edition ed., Hemisphere Publishing Corporation, New York.

Huang, Y., Paul, D. R., 2006. Physical Aging of Thin Glassy Polymer Films Monitored by Optical Properties. Macromolecules 39, 1554-1559.

Joshi, Y. M. 2007. Model for cage formation in colloidal suspension of laponite. J. Chem. Phys. 127, 081102.

Joshi Y. M., Reddy G. R. K., Kulkarni A. L., Kumar N., Chhabra R. P.; 2008 Rheological Behavior of Aqueous Suspensions of Laponite: New Insights into the Ageing Phenomena. Proc. Roy. Soc. A 464, 469-489.

Joshi Y. M. and Reddy, G. R. K., 2007 Aging in a colloidal glass in creep flow: A master curve for creep compliance. *arXiv:0710.5264*.

Lagaly, G., 1989. Principles of flow of kaolin and bentonite dispersions. Appl. Clay Sci. 4, 105-123.

Lorentz, H. A. 1952. The theory of electrons. Dover, New York.

Mossa, S., De Michele, C., Sciortino, F., 2007. Aging in a Laponite colloidal suspension: A Brownian dynamics simulation study. J. Chem. Phys. 126, 014905.

Mourchid, A., Delville, A., Lambard, J., Lecolier, E., Levitz, P., 1995. Phase Diagram of Colloidal Dispersions of Anisotropic Charged Particles: Equilibrium Properties, Structure, and Rheology of Laponite Suspensions. Langmuir 11, 1942-1950.

Mourchid, A., Levitz, P., 1998. Long-term gelation of laponite aqueous dispersions. Phys. Rev. E 57, R4887-R4890.

Niskanen, I., Raty, J., Peiponen, K. E., 2006. A multifunction spectrophotometer for measurement of optical properties of transparent and turbid liquids. Measurement Sci. & Tech. 17, N87-N91.

Ramsay, J. D. F., 1986. Colloidal Properties of Synthetic Hectorite Clay Dispersions. J. Colloid Interface Sci. 109, 441-447.

Ruzicka, B., Zulian, L., Ruocco, G., 2004. Routes to Gelation in a Clay Suspension. Phys. Rev. Lett. 93, 258301.

Ruzicka, B., Zulian, L., Ruocco, G., 2006. More on the Phase Diagram of Laponite. Langmuir 22, 1106-1111.

Schiebener, P., Straub, J., Sengers, J., Gallagher, J. S., 1990. Refractive index of water and steam as a function of wavelength, temperature and density. J. Phys. and Chem. Ref. Data 19, 677-717.

Schosseler, F., Kaloun, S., Skouri, M., Munch, J.P., 2006. Diagram of the aging dynamics in laponite suspensions at low ionic strength. Phys. Rev. E 73, 021401.

Tan, C. Z., 1998. Review and analysis of refractive index temperature dependence in amorphous SiO2. J. Non-Crystalline Solids 238, 30-36.

Tanaka, H., Jabbari-Farouji, S., Meunier, J., Bonn, D., 2005. Kinetics of ergodic-to-nonergodic transitions in charged colloidal suspensions: Aging and gelation. Phys. Rev. E 71, 021402.

Tanaka, H., Meunier, J., Bonn, D., 2004. Nonergodic states of charged colloidal suspensions: Repulsive and attractive glasses and gels. Phys. Rev. E 69, 031404.

Tanio, N., Nakanishi, T., 2006. Physical Aging and Refractive Index of Poly(methyl methacrylate) Glass. Polymer Journal 38, 814-818.

Van Olphen, H., 1977. An Introduction to Clay Colloid Chemistry. Wiley New York.

Wu, S. T., 1986. Birefringence dispersions of liquid crystals. Phys. Rev. A 33, 1270-1274.

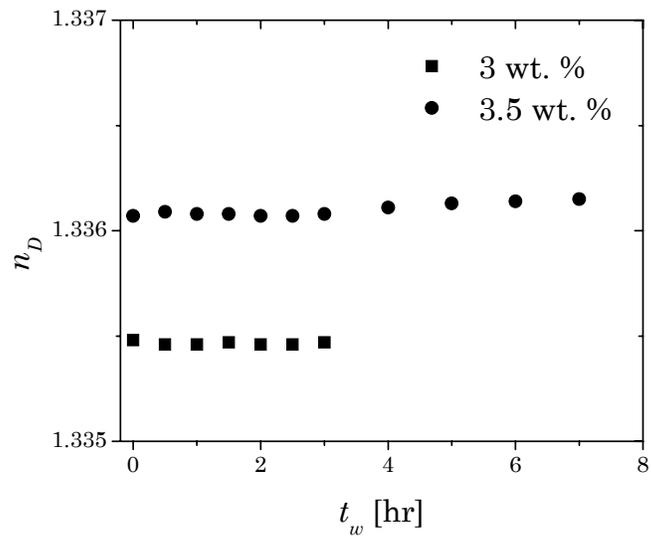

**Figure 1.** Refractive index of 3 and 3.5 weight % Laponite dispersion as a function of ageing time at 20°C.

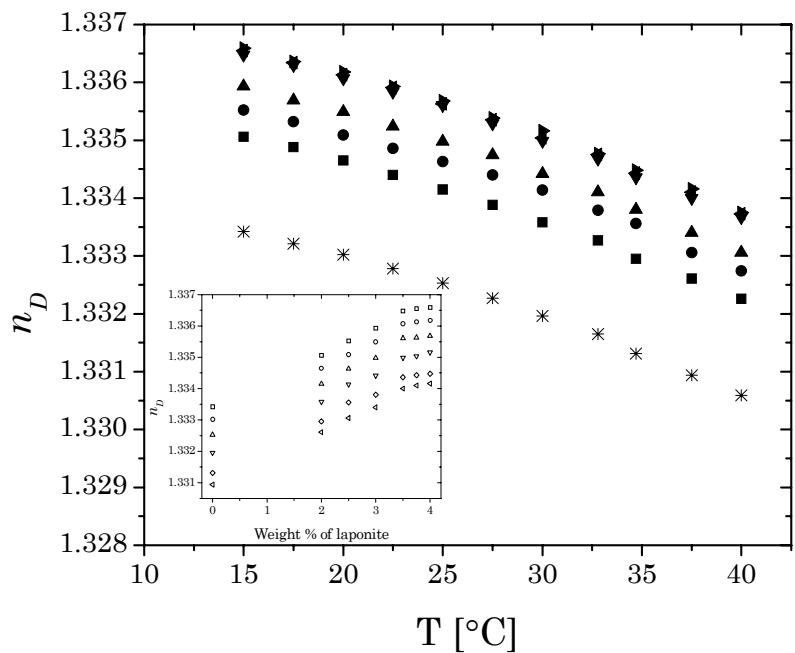

**Figure 2.** Refractive index of aqueous Laponite dispersions as a function of temperature for various concentrations of Laponite. From top to bottom, concentration weight % considered are: 4, 3.75, 3.5, 3, 2.5, 2, 0 (pure water). Inset shows dependence of RI on concentration for various temperatures (from top to bottom, 15, 20, 25, 30, 35 and 40°C).

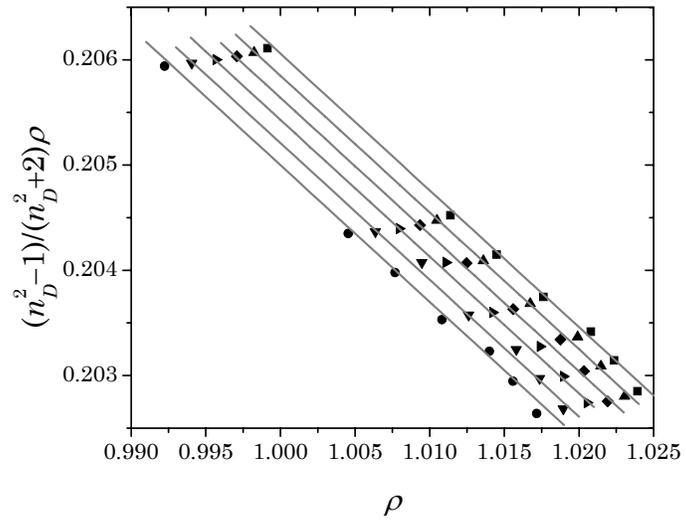

**Figure 3.** Fit of the extended Lorenz-Lorentz equation (Equation 2) at various concentrations and temperatures of the Laponite dispersion in water. From left to right, temperatures considered are 40, 35, 30, 25, 20 and 15 °C.